# Fluctuation-Electromagnetic Interaction: Effects of Uniform Motion, Rotation, and Thermal Disequilibrium

G.V. Dedkov and A.A. Kyasov

Nanoscale Physics Group, Kabardino-Balkarian State University, Nalchik, 360004, Russia

In this work, we give a consistent review of recent analytical results of reference character related to the fluctuation-electromagnetic interactions in the systems particle –vacuum, particle –surface, particle –particle, and surface –surface. Effects of dynamical, rotational, and thermal disequilibrium are considered. The applications of these results are important in atomic and molecular physics, quantum field theory and nanotechnology.

## 1. Introduction

The problems of interaction at the nanonoscale, such as attraction (repulsion) forces, friction and heating effects, etc. are among the oldest problems in physics and chemistry, and among the most important in emerging technology of nanomachines. What has been intensively debated is a microscopic origin of friction and dissipation in a system of the bodies in relative motion, interactions with the absence thermal and dynamic equilibrium, the role of material and geometric factors, etc. Significant role in this case is played by the interactions mediated by fluctuation electromagnetic field which is ubiquitous.

So, electromagnetic fluctuation-induced forces between atoms and surfaces are generally known as Casimir-Polder interactions, and between the surfaces---as Van der Waals-Casimir-Lifshitz interactions. Friction force arising in vacuum contact between electrically neutral bodies in relative motion can be thought of as an elementary force of dissipative nature, that still can be calculated from first principles. Also, radiative heating or cooling of a body in the thermal field of another body is the manifestation of radiation heat transfer mediated by evanescent and propagating electromagnetic waves .

Despite the overall physical origin of the above-mentioned effects, the problems of calculating conservative / dissipative forces and heat transfer in the static/dynamic and different geometric configurations, as well as in the systems with/without thermal equilibrium are often treated separately from each other in the framework of various theoretical approaches (see [1-8] for a review and numerous references therein). So, some authors preferred using the configuration of two plates---a classical configuration by Casimir and Lifshitz, that resulted in erroneous results upon the dynamic modification of the static Lifshitz's theory in configuration particle-surface [7].

In contrast to this, in our works [9-11] we have developed general approach to these problems within the formalism of fluctuation electrodynamics [12], with particular attention devoted to the particle-vacuum, particle-surface, and particle-particle configurations. These basic results were recently recovered and confirmed within the fully covariant calculation of radiation forces on a polarizable particle moving in free vacuum and above a plane surface [6,13]. Since the co-moving reference frame is a natural candidate for local thermodynamic equilibrium, from the view point of relativistic thermodynamics the two situations of bodies in relative motion or fixed at different temperature represent very similar non-equilibrium settings. As we will see in what follows, this unity is fundamental since the temperature of a moving body in the co-moving reference frame and the Doppler-shifted frequency are combined into a single variable. Moreover, the concept of local thermal equilibrium is a prerequisite to apply the fluctuation-dissipation theorem in the relativistic concept [6,13].

The present review has the following structure. In Sec. 2 we give a list of the main symbols used. Section 3 is devoted to a short description of the calculation method, summarizing the main results and references in Table 1. Section 4 contains a set of analytical expressions for the conservative/dissipative fluctuation-electromagnetic (radiation) forces and the rates of heat exchange in configurations of interest, and in Sec. 5 we give concluding remarks.

## 2. List of symbols used

$V = |\mathbf{V}|$ – velocity of a particle; $\beta = V/c$; $c$ – speed of light in vacuum;

$\gamma = (1-\beta^2)^{-1/2}$ – Lorentz-factor;

$k_B$ and $\hbar$ – Boltzmann's and Planck's constants;

$l$ – separation distance between the two plates; $S$ – surface area of the plates;

$z_0, z$ – separation distance;

$D'_{ik}(\omega, \mathbf{R})$, $D''_{ik}(\omega, \mathbf{R})$ --real and imaginary parts of the Green function $D_{ik}(\omega, \mathbf{R})$; $i,k = x, y, z$

$R$ – particle radius, interparticle separation;

$F_z$ – normal force;

$F_x$ – lateral (tangential) force;

$M_x, M_z$ – components of torque;

$U(z_0), U(R)$ – interaction potential, free energy;

$\dot{Q}$ – heating (cooling) rate;

$T$ – equilibrium temperature;

$T_1$ – temperature of the moving body in the co-moving reference frame;

$T_2$ – temperature of the resting body;

$\omega$ – frequency; $\omega_0$ – characteristic absorption frequency;

$\Omega$ – angular rotation velocity;

$\omega^{\pm} = \omega_{\pm} = \omega \pm \Omega; \omega^{\pm} = \omega_{\pm} = \omega \pm k_x V$;

$\varepsilon, \varepsilon(\omega), \varepsilon_1(\omega), \varepsilon_2(\omega)$ – dielectric permittivity;

$\mu, \mu(\omega), \mu_1(\omega), \mu_2(\omega)$ – magnetic permittivity;

$\varepsilon', \varepsilon''$ – real and imaginary parts of the frequency-dependent dielectric permittivity;

$\alpha(\omega), \alpha_1(\omega), \alpha_2(\omega), \alpha_e(\omega)$ – dipole electric polarizability;

$\alpha'(\omega), \alpha''(\omega)$ – real and imaginary parts of $\alpha(\omega)$;

$\alpha_m(\omega), \alpha'_m(\omega), \alpha''_m(\omega)$ – magnetic polarizability, real and imaginary parts of $\alpha_m(\omega)$;

$\Delta(\omega) = (\varepsilon(\omega) - 1)/(\varepsilon(\omega) + 1); \Delta'(\omega)$ and $\Delta''(\omega)$ are the real and imaginary parts;

$\rho$ – local charge density;

$\mathbf{E}, \mathbf{B}, \mathbf{H}$ – electric field, magnetic induction and magnetic field strength;

$\mathbf{P}, \mathbf{M}$ – polarization and magnetization of continuous medium;

$\mathbf{d}, \mathbf{m}$ – dipole electric and magnetic moments;

$\mathbf{k}$ – wave-vector, $|\mathbf{k}| = k$; $k_x, k_y$ – projections of $\mathbf{k}$;

$\mathbf{j}$ – local current density.

## 3. General relations and references

The essence of our calculation method can be demonstrated in the case of two objects, one of which is a small uniformly moving (rotating) particle, and another one is the resting body. The latter can be another particle, the extended body (thick plate) or a vacuum background. Correspondingly, we use two reference coordinate systems, namely $\Sigma$ and $\Sigma'$ related to the moving and resting bodies. This allows us to take advantages of the relativistic invariance in consideration of the problem and this attaches particular importance to the given configuration. Without loss of generality, let us consider the configuration shown in Fig. 1. The particle undergoes an action of fluctuation electromagnetic field created by the photon gas (of definite temperature) and the heated thick plate. Components $\mathbf{E}$ and $\mathbf{B}$ of the field should obey the Maxwell equations with account of necessary boundary conditions. Denoting $\rho$ and $\mathbf{j}$ to be the local charge and current densities, the Lorentz force applied to the particle is given by

$$\mathbf{F} = \int \langle \rho \mathbf{E} \rangle d^3 r + \frac{1}{c} \int \langle \mathbf{j} \times \mathbf{B} \rangle d^3 r \qquad (1)$$

where the integrations are performed over the volume of the particle and $\langle ... \rangle$ denotes total quantum-statistical averaging. According to the basic identities from electrodynamics,

$$\rho = -\nabla \mathbf{P}, \quad \mathbf{j} = \partial \mathbf{P}/\partial t + c \cdot \nabla \times \mathbf{M} \qquad (2)$$

where $\mathbf{P}$ and $\mathbf{M}$ denote the polarization and magnetization vectors. Note that all variables in (1),(2) are given in the resting ("laboratory") system $\Sigma$. Assuming the case shown in Fig. 1 and the condition of dipole approximation $z_0/R \gg 1$, vectors $\mathbf{P}$ and $\mathbf{M}$ of a moving particle are

$$\mathbf{P}(\mathbf{r},t) = \mathbf{d}(t)\delta(\mathbf{r} - \mathbf{V} \cdot t), \mathbf{M}(\mathbf{r},t) = \mathbf{m}(t)\delta(\mathbf{r} - \mathbf{V} \cdot t) \qquad (3)$$

Using the Maxwell equations $\nabla \times \mathbf{E} = -(1/c)\partial \mathbf{B}/\partial t, \nabla \mathbf{B} = 0$ and (2), (3), and performing integration in (1) yields [10]

$$\mathbf{F} = \langle \nabla(\mathbf{d} \cdot \mathbf{E} + \mathbf{m} \cdot \mathbf{B}) \rangle \qquad (4)$$

It is worth noting that in the case where the dipole moments and electromagnetic field have regular character, Eq. (4) contains the additional term $\frac{1}{c}\left\langle \frac{d}{dt}(\mathbf{d} \times \mathbf{B}) \right\rangle$ in the right hand side [10].

To calculate the heating rate of the particle, we first consider this quantity in the particle rest frame $\Sigma'$, where it is obviously expressed through the dissipation integral

$$\frac{dQ'}{dt'} = \int \langle \mathbf{j'} \cdot \mathbf{E'} \rangle d^3 r' = \gamma^2 \left( \int \langle \mathbf{j} \cdot \mathbf{E} \rangle d^3 r - \mathbf{F} \cdot \mathbf{V} \right) \qquad (5)$$

where the second (right) part of Eq. (5) is obtained using relativistic transformations for the current density, electric field and volume. Since $dQ'/dt' = \gamma^2 dQ/dt$ according to Planck's formulation of relativistic thermodynamics, Eq. (5) after calculating the dissipation integral takes a more compact form [10]

$$dQ/dt = \dot{Q} = \langle \dot{\mathbf{d}} \cdot \mathbf{E} + \dot{\mathbf{m}} \cdot \mathbf{B} \rangle \qquad (6)$$

Finally, according to our calculation method [9-11,17], the quantities in the right hand sides of (4) and (6) are represented as the sums of binary products of spontaneous and induced quantities

$$\mathbf{F} = \left\langle \nabla \left( \mathbf{d}^{sp} \cdot \mathbf{E}^{ind} + \mathbf{d}^{ind} \cdot \mathbf{E}^{sp} + \mathbf{m}^{sp} \cdot \mathbf{B}^{ind} + \mathbf{m}^{ind} \cdot \mathbf{B}^{sp} \right) \right\rangle \tag{7}$$

$$dQ/dt = \left\langle \dot{\mathbf{d}}^{sp} \cdot \mathbf{E}^{ind} + \dot{\mathbf{d}}^{ind} \cdot \mathbf{E}^{sp} + \dot{\mathbf{m}}^{sp} \mathbf{B}^{ind} + \dot{\mathbf{m}}^{ind} \mathbf{B}^{sp} \right\rangle \tag{8}$$

Equations (7) and (8) are the basic starting equations in configurations particle-vacuum (Fig. 1) and particle-surface (Fig. 2).

In the case of interaction between the two polarizable particles in vacuum, of which the first rotates with the angular velocity $\Omega$ (Fig. 3), the equations that are analogous to (4), (6) can be conveniently written in the form [14]

$$U(R) = -\frac{1}{2} \left\langle \mathbf{d}_1(t) \mathbf{E}(\mathbf{r}_1,t) \right\rangle \tag{9}$$

$$\dot{Q} = \left\langle \dot{\mathbf{d}}_1(t) \mathbf{E}(\mathbf{r}_1,t) \right\rangle \tag{10}$$

$$M_{x,y,z} = \left\langle \mathbf{d}_1(t) \times \mathbf{E}(\mathbf{r}_1,t) \right\rangle_{x,y,z} \tag{11}$$

where $U(R)$ and $M_{x,y,z}$ denote the free energy of the system and the torque on the first (rotating) particle, while all the quantities correspond to the location point of this particle and refer to the system $\Sigma$. In the same manner, the right hand sides of (9)—(11) should be separated onto a sum of the binary products of spontaneous and induced components, while the former ones must include the spontaneous moments of both particles and the spontaneous field of vacuum.

Further calculations with the use of Eqs. (7)—(11) are performed using the fluctuation-dissipation relations for the dipole moments and electromagnetic fields that are written in the rest frame $\Sigma'$ of the moving (rotating) particle and reference frames $\Sigma$ of vacuum, planar interface, or a non-rotating particle. Despite a tedious algebra, the calculations are straightforward (see [9-11,14-18] in more detail). Table 1 below summarizes a short description of the main results with reference to Sec. 4, as well as the corresponding references from literature.

**Table 1 Basic results and references**

| Configuration | Conditions | Physical value | Equation | Ref. |
|---|---|---|---|---|
| particle-vacuum | $V \to c, T_1 \neq T_2$ (r) <br> $V \ll c, T_1 \neq T_2$ <br> $V \ll c, T_1 = T_2 = T$ | $F_x, \dot{Q}$ <br> $F_x, Q$ <br> $F_x$ | (12),(13) <br> (14) <br> (15) | [11] |
| particle-surface | $V \to c, T_1 \neq T_2 = T_3$ (r) <br> $V \ll c, T_1 \neq T_2$ (nr) <br> $V \ll c, T_1 = T_2 = T$ (nr) | $F_x, F_z, \dot{Q}$ <br> $F_x, F_z, \dot{Q}$ <br> $F_x, F_z$ | (16),(17),(18) <br> (22),(23),(24) <br> (25),(26) | [10],[11] |
| atom-metallic surface | $V \ll c, T_1 = T_2 = T_3 = 0$ (nr) <br> $V \to c, T_1 = T_2 = T_3 = 0$ (r) <br> $\Delta_e(\omega) = 1, \Delta_m(\omega) = -1$ | $F_z$ <br> $F_z$ | (30) <br> (31) | [15] |
| particle-surface | $V = 0, T_1 = 0,$ <br> $T_2 = T_3 = T$ (r) | $F_z$ | (32) | [17] |
| atom-surface | $V = 0, T_1 = 0, T_2 \neq T_3$ (r) <br> no absorption and scattering by an atom | $F_z$ | (33),(34) | [17] |
| particle-surface | Eq.(35), $T_1 \neq T_2$, (nr) <br> $\boldsymbol{\Omega} = (0,0,\Omega), R \ll z_0$ | $F_z, \dot{Q}, M_z$ | (36),(37),(38) | [16] |
| particle-surface | Eq.(35), $T_1 \neq T_2$, (nr) <br> $\boldsymbol{\Omega} = (\Omega,0,0), R \ll z_0$ | $F_z, \dot{Q}, M_x$ | (39),(40),(41) | [18] |
| particle-particle | $\boldsymbol{\Omega} = (0,0,\Omega), a_{1,2} \ll R$ (r) <br> $T_1 \neq T_2, \neq T_3$ | $F_z, \dot{Q}, M_z$ | (42),(43),(44) | [14] |
| particle-particle | $\boldsymbol{\Omega} = (0,0,\Omega)$ (nr) <br> $c \to \infty, a_{1,2} \ll R$ <br> $T_1 \neq T_2, \neq T_3$ | $F_z, \dot{Q}, M_z$ | (46),(47),(48) | [14] |
| particle-particle | $a_{1,2} \ll R, \Omega = 0$ (r) <br> $T_1 \neq T_2, \neq T_3$ <br> $a_{1,2} \ll R, \Omega = 0$ (r) <br> $T_1 = T_2, = T_3 = T$ <br> $a_{1,2} \ll R, \Omega = 0$ (r) <br> $T_1 = T_2, = T_3 = 0$ | $F_z$ <br> $F_z$ <br> $F_z$ | (50) <br> (51) <br> (52) | [14] |
| surface-surface | $V \ll c,$ <br> $T_1 = T_2 = T_3 = T$ (r) | $F_x, F_z, \dot{Q}$ | (52),(53),(54) | [10] |
| Surface-surface | $V \ll c, T_1 \neq T_2$ (nr) | $F_x, F_z, \dot{Q}$ | (55),(56),(57) | [10] |

Note: "Reference" column contains only our works; the works of other authors are referred to through the text; (r) –retarded interaction; (nr) –nonretarded interaction

## 4. Basic reference results

### 4.1 A uniformly moving particle in a vacuum background

Figure 1 shows the geometrical configuration and coordinate systems used.

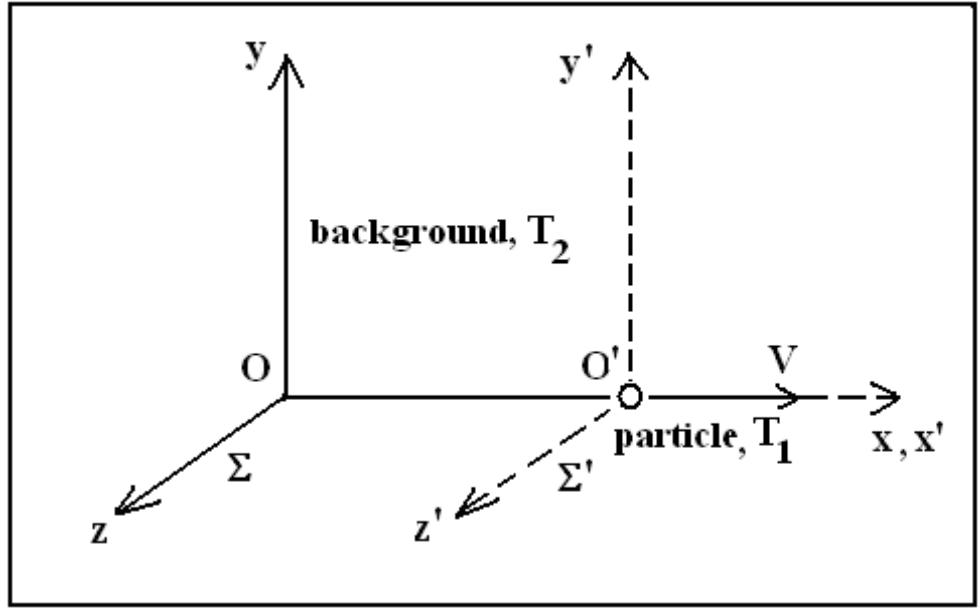

Fig.1 Cartesian reference frame $\Sigma$ associated with the vacuum background (in rest) and the co-moving reference frame of the particle $\Sigma'$.

The tangential force $F_x$ and heating rate $\dot{Q}$ are given by [11]

$$F_x = -\frac{\gamma \hbar}{\pi c^4} \int_0^\infty d\omega\, \omega^4 \int_{-1}^{1} dx\, x(1+\beta x)^2 \cdot [\alpha_e''(\gamma\omega(1+\beta x)) + \alpha_m''(\gamma\omega(1+\beta x))] \cdot$$
$$\cdot \left[ \coth\left(\frac{\hbar\omega}{2k_B T_2}\right) - \coth\left(\frac{\gamma\hbar\omega(1+\beta x)}{2k_B T_1}\right) \right] \quad (12)$$

$$\dot{Q} = \frac{\gamma\hbar}{\pi c^3} \int_0^\infty d\omega\, \omega^4 \int_{-1}^{1} dx\, (1+\beta x)^3 \cdot [\alpha_e''(\gamma\omega(1+\beta x)) + \alpha_m''(\gamma\omega(1+\beta x))] \cdot$$
$$\cdot \left[ \coth\left(\frac{\hbar\omega}{2k_B T_2}\right) - \coth\left(\frac{\gamma\hbar\omega(1+\beta x)}{2k_B T_1}\right) \right] \quad (13)$$

At $\beta \ll 1$, Eq. (12) reduces to

$$F_x = -\frac{4\hbar V}{3\pi c^5} \int_0^\infty d\omega\, \omega^5 \cdot \left\{ \begin{array}{l} \dfrac{\hbar}{4k_B T_1} \dfrac{\alpha_e''(\omega) + \alpha_m''(\omega)}{\sinh^2(\hbar\omega/2k_B T_1)} + 2[\Pi(\omega, T_2) - \Pi(\omega, T_1)] \cdot \\ \cdot \left[ \dfrac{\alpha_e''(\omega) + \alpha_m''(\omega)}{\omega} + \dfrac{1}{2}\dfrac{d\alpha_e''(\omega)}{d\omega} + \dfrac{1}{2}\dfrac{d\alpha_m''(\omega)}{d\omega} \right] \end{array} \right\} \quad (14)$$

Particularly, at $T_1 = T_2 = T$, Eq.(14) takes the form [11,19]

$$F_x = -\frac{\hbar^2}{3\pi c^4}\frac{\beta}{k_B T}\int_0^\infty d\omega \omega^5 [\alpha_e''(\omega) + \alpha_m''(\omega)]\sinh^{-2}(\hbar\omega/2k_B T) \qquad (15)$$

### 4.2 A particle moving above a plane surface

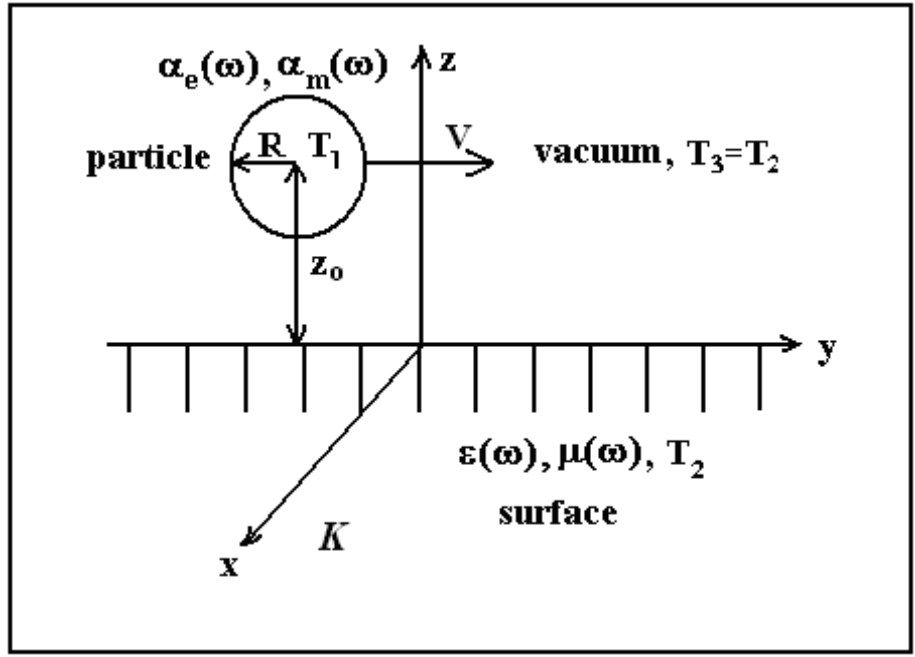

Fig. 2. The geometrical configuration and Cartesian reference frame of the surface and vacuum background (in rest) $\Sigma$, assuming mutual thermal equilibrium at temperature $T_2$. The co-moving reference frame $\Sigma'$ of a moving particle is not shown

General relativistic expressions for the components of tangential and normal force and the heating rate of a particle in the reference frame $\Sigma$ are given by [10]

$$F_x = -\frac{\hbar\gamma}{2\pi^2}\int_0^\infty d\omega \int_{-\infty}^{+\infty} dk_x \int_{-\infty}^{+\infty} dk_y k_x \left[\alpha_e''(\gamma\omega^+)\mathrm{Im}\left(\frac{\exp(-2q_0 z)}{q_0}R_e(\omega,\mathbf{k})\right) + (e\leftrightarrow m)\right] \cdot$$
$$\cdot \left[\coth\left(\frac{\hbar\omega}{2k_B T_2}\right) - \coth\left(\frac{\gamma\hbar\omega^+}{2k_B T_1}\right)\right] + (\ldots) \qquad (16)$$

$$F_z = -\frac{\hbar\gamma}{2\pi^2}\int_0^\infty d\omega \int_{-\infty}^{+\infty} dk_x \int_{-\infty}^{+\infty} dk_y \left\{\begin{array}{l}\alpha_e''(\gamma\omega^+)\mathrm{Re}[\exp(-2q_0 z)R_e(\omega,\mathbf{k})]\coth\left(\frac{\gamma\hbar\omega^+}{2k_B T_1}\right) + \\ + \alpha_e'(\gamma\omega^+)\mathrm{Im}[\exp(-2q_0 z)R_e(\omega,\mathbf{k})]\coth\left(\frac{\hbar\omega}{2k_B T_2}\right) + (e\leftrightarrow m)\end{array}\right\} \qquad (17)$$

$$\frac{dQ}{dt} = \frac{\hbar\gamma}{2\pi^2}\int_0^\infty d\omega \int_{-\infty}^{+\infty} dk_x \int_{-\infty}^{+\infty} dk_y \omega^+ \cdot \left[\alpha''_e(\gamma\omega^+)\text{Im}\left(\frac{\exp(-2q_0 z)}{q_0}R_e(\omega,\mathbf{k})\right) + (e \leftrightarrow m)\right]$$
$$\cdot \left[\coth\left(\frac{\hbar\omega}{2k_B T_2}\right) - \coth\left(\frac{\gamma\hbar\omega^+}{2k_B T_1}\right)\right] + (...) \quad (18)$$

where $\omega^+ = \omega + k_x V$,

$$R_e(\omega,\mathbf{k}) = \Delta_e(\omega)\left[2(k^2 - k_x^2\beta^2)(1 - \omega^2/k^2 c^2) + (\omega^+)^2/c^2\right] +$$
$$+ \Delta_m(\omega)\left[2k_y^2\beta^2(1 - \omega^2/k^2 c^2) + (\omega^+)^2/c^2\right] \quad (19)$$

$$R_m(\omega,\mathbf{k}) = \Delta_m(\omega)\left[2(k^2 - k_x^2\beta^2)(1 - \omega^2/k^2 c^2) + (\omega^+)^2/c^2\right] +$$
$$+ \Delta_e(\omega)\left[2k_y^2\beta^2(1 - \omega^2/k^2 c^2) + (\omega^+)^2/c^2\right] \quad (20)$$

$$\Delta_e(\omega) = \frac{q_0\varepsilon(\omega) - q}{q_0\varepsilon(\omega) + q}, \quad \Delta_m(\omega) = \frac{q_0\mu(\omega) - q}{q_0\mu(\omega) + q}, \quad q = \left(k^2 - (\omega^2/c^2)\varepsilon(\omega)\mu(\omega)\right)^{1/2},$$
$$q_0 = (k^2 - \omega^2/c^2)^{1/2}, \quad k^2 = k_x^2 + k_y^2 \quad (21)$$

The terms (…) are identical to Eqs. (12) and (13) corresponding to the configuration particle – vacuum [11].

In the electrostatic (nonrelativistic) limit , $c \to \infty$, Eqs. (16)—(18) take the form

$$F_x = -\frac{\hbar}{\pi^2}\int_0^\infty d\omega \int_{-\infty}^{+\infty} dk_x \int_{-\infty}^{+\infty} dk_y k k_x \exp(-2kz)\Delta''(\omega)\alpha''_e(\omega^+)\left[\coth\left(\frac{\hbar\omega}{2k_B T_2}\right) - \coth\left(\frac{\hbar\omega^+}{2k_B T_1}\right)\right] \quad (22)$$

$$F_z = -\frac{\hbar}{\pi^2}\int_0^\infty d\omega \int_{-\infty}^{+\infty} dk_x \int_{-\infty}^{+\infty} dk_y k^2 \exp(-2kz) \cdot$$
$$\cdot \left[\Delta''(\omega)\alpha'_e(\omega^+)\coth\left(\frac{\hbar\omega}{2k_B T_2}\right) + \Delta'(\omega)\alpha''_e(\omega^+)\coth\left(\frac{\hbar\omega^+}{2k_B T_1}\right)\right] \quad (23)$$

$$\frac{dQ}{dt} = \frac{\hbar}{\pi^2}\int_0^\infty d\omega \int_{-\infty}^{+\infty} dk_x \int_{-\infty}^{+\infty} dk_y k \exp(-2kz)\Delta''(\omega)\alpha''_e(\omega^+)\omega^+\left[\coth\left(\frac{\hbar\omega}{2k_B T_2}\right) - \coth\left(\frac{\hbar\omega^+}{2k_B T_1}\right)\right] \quad (24)$$

There are also some special cases of these equations.

At $T_1 = T_2 = T$, from (22) we obtain [20]

$$F_x = \frac{3}{2\pi}\frac{\hbar V}{z^5}\int_0^\infty d\omega\, \alpha''(\omega)\Delta''(\omega)\frac{d}{d\omega}\frac{1}{\exp(\hbar\omega/k_B T) - 1} \quad (25)$$

In the limiting case $T_1 = T_2 = T \to 0$, Eq. (22) reduces to [18]:

$$F_x = -\frac{4\hbar V}{\pi^2 z^5} \int_0^\infty du (u^3 K_0(2u) + 0.5 u^2 K_1(2u)) \int_0^u dp\, \alpha''(\omega_0(u-p))\Delta''(\omega_0 p),\ \omega_0 = V/z \tag{26}$$

where $K_0(x)$ are $K_1(x)$ – are the McDonald functions. On the other hand (at $T_1 = T_2 = T \to 0$), Eq.(23) reduces to [15]

$$F_z = -\frac{\hbar}{\pi^2} \int_{-\infty}^{+\infty} dk_x \int_{-\infty}^{+\infty} dk_y\, k^2 \exp(-2k z_0) \cdot \mathrm{Im}\left[i \int_0^\infty d\xi\, \Delta(i\xi)\alpha(i\xi + k_x V)\right] +$$
$$+ \frac{4\hbar}{\pi^2} \int_0^\infty dk_x \int_0^\infty dk_y\, k^2 \exp(-2k z_0) \int_0^{k_x V} d\omega\, \Delta'(\omega)\alpha''(\omega - k_x V), \tag{27}$$

Using a non-dissipative model of metallic half –space,

$$\varepsilon(\omega) = 1 - \omega_p^2/\omega^2,\ \Delta(i\xi) = \frac{\omega_s^2}{\omega_s^2 + \xi^2},\ \omega_s = \omega_p/\sqrt{2}, \tag{28}$$

and a single-oscillator model of the atomic polarizability

$$\alpha(i\xi) = \frac{\alpha(0)\omega_0^2}{\omega_0^2 + \xi^2} \tag{29}$$

where $\omega_p$ is the plasma frequency, $\alpha(0)$ is the static value of the dipole polarizability, and $\omega_0$ is the atomic transition frequency, Eq. (27) is reduced to the form

$$F_z = \frac{\hbar \omega_s \alpha(0)}{8\pi z^4} \int_0^\infty x^3 \frac{d^3 K_0(x)}{dx^3} \left[\frac{(1+\eta)\theta(1-qx)}{(1+\eta)^2 - q^2 x^2} + \frac{\theta(qx-1)}{[1-(\eta+qx)^2]}\right] dx +$$
$$+ \frac{\hbar \omega_s \alpha(0)}{8\pi z^4} \eta^2 \int_{1/q}^\infty \frac{d^3 K_0(x)}{dx^3} \frac{x^3}{\eta^2 - (1-qx)^2} dx,\ \eta = \omega_s/\omega_0,\ \omega_s = \omega_p/\sqrt{2} \tag{30}$$

where $\theta(x)$ is the Heavyside step –function.

Relativistic analog of Eq. (16) was obtained in [15], as well. We omit the general expression for brevity. In the case of an ideally conducting cavity wall, $\varepsilon(\omega) \to \infty$, it follows $\Delta_e(\omega) = 1$, $\Delta_m(\omega) = -1$, and Eq. (16) takes the form

$$F_z = \frac{4\hbar \alpha(0)\omega_0^5}{\pi^2 c^4 \gamma^6} \int_0^\infty dx \int_0^\infty dy\, \frac{(1 - \beta^2 x^2 + y^2)(x^2 + y^2)^{3/2}}{[(1+\beta x)^2 + y^2][(1-\beta x)^2 + y^2]} \left[\frac{d^3 K_0(t)}{dt^3}\right]_{t=\lambda\sqrt{x^2+y^2}} +$$
$$+ \frac{2\hbar \alpha(0)\omega_0^5}{\pi c^4 \gamma^6} \int_{1/\beta}^\infty dx \left[x^2(1-\beta^2) + 2\beta x - 1\right]^{3/2} \left[\frac{d^3 K_0(t)}{dt^3}\right]_{t=\lambda\sqrt{x^2(1-\beta^2)+2\beta x-1}} \tag{31}$$

where $\lambda = 2\omega_0 z_0 / \gamma c$.

There are also several interesting static situations out of thermal equilibrium. So, in configuration particle –surface ($V = 0, T_1 = 0, T_2 = T_3 = T$) from (17) we obtain [17]

$$F_z = -2k_B T \sum_{n=0}^{\infty} a_n \int_0^{\infty} dk k [R_e(i\xi_n, k)\alpha_e(i\xi_n) + R_m(i\xi_n, k)\alpha_m(i\xi_n)] \exp\left(-2\sqrt{k^2 + \xi_n^2/c^2}\, z\right) +$$
$$+ \frac{2\hbar}{\pi} \int_0^{\infty} d\omega \Pi(\omega, T) \alpha_e''(\omega) \mathrm{Re}\left\{\int_0^{\infty} dk k \exp(-2q_0 z) R_e(\omega, k)\right\} + [\alpha_e'' \to \alpha_m'', R_e \to R_m], \quad (32)$$

where the term $[...]$ is the same as the second integral with the following replacements:

$a_n = (1 - \delta_{0n}/2)$, $\xi_n = 2\pi k_B T n/\hbar$, $\Pi(\omega, T) = (\exp(\hbar\omega/k_B T) - 1)^{-1}$.

The case $V = 0, T_1 = 0, T_2 \neq T_3$ in configuration particle-surface (atom-surface) has been firstly considered in [2], assuming that the processes of scattering and absorption of radiation on an atom could be neglected. The resulting formula can be also written in an identical form [17]

$$F_z = -2k_B T_2 \sum_{n=0}^{\infty} a_n \int_0^{\infty} dk k [R_e(i\xi_n, k)\alpha_e(i\xi_n) + R_m(i\xi_n, k)\alpha_m(i\xi_n)] \exp\left(-2\sqrt{k^2 + \xi_n^2/c^2}\, z\right) +$$
$$+ F_{th}(T_2, 0, z) - F_{th}(T_3, 0, z), \xi_n = 2\pi k_B T_2 n/\hbar \quad (33)$$

$$F_{th}(T, 0, z) = -\frac{2\hbar}{\pi} \int_0^{\infty} d\omega \alpha_e'(\omega) \Pi(\omega, T)\left\{\int_{\omega/c}^{\infty} dk k \,\mathrm{Im} R_e \exp(-2q_0 z)\right\} + [\alpha_e' \to \alpha_m', R_e \to R_m] \quad (34)$$

The second term in (34) has the same form as the first one with the corresponding replacements. Another representation of Eqs. (33), (34) is given in [2]

### 4.3 A particle rotating near a surface

Figure 3 shows a configuration in which the axis of rotation of the particle is directed along the normal to the surface [16]. Assuming that the particle is a point-like nonrelativistic dipole, the following conditions must be fulfilled

$$R << \min\left\{\frac{2\pi c}{\omega_0}, \frac{2\pi c}{\Omega}, \frac{2\pi \hbar c}{k_B T_1}, \frac{2\pi \hbar c}{k_B T_2}\right\}, R << z_0 \quad (35)$$

$$F_z = -\frac{3\hbar}{16\pi z_0^4} \int_{-\infty}^{+\infty} d\omega \left\{ \begin{array}{l} \Delta'(\omega)\alpha''(\omega)\coth\frac{\hbar\omega}{2k_BT_1} + \Delta''(\omega)\alpha'(\omega)\coth\frac{\hbar\omega}{2k_BT_2} + \\ + \Delta'(\omega)\alpha''(\omega^+)\coth\frac{\hbar\omega^+}{2k_BT_1} + \Delta''(\omega)\alpha'(\omega^+)\coth\frac{\hbar\omega}{2k_BT_2} \end{array} \right\} \quad (36)$$

$$\dot{Q} = \frac{\hbar}{8\pi z_0^3} \int_{-\infty}^{+\infty} d\omega\,\omega \left\{ \begin{array}{l} \Delta''(\omega)\alpha''(\omega)\left[\coth\frac{\hbar\omega}{2k_BT_2} - \coth\frac{\hbar\omega}{2k_BT_1}\right] + \\ \Delta''(\omega)\alpha''(\omega^+)\left[\coth\frac{\hbar\omega}{2k_BT_2} - \coth\frac{\hbar\omega^+}{2k_BT_1}\right] \end{array} \right\} \quad (37)$$

$$M_z = \frac{\hbar}{4\pi z_0^3} \int_{-\infty}^{+\infty} d\omega\,\Delta''(\omega)\alpha''(\omega^+) \cdot \left[\coth\frac{\hbar\omega^+}{2k_BT_1} - \coth\frac{\hbar\omega}{2k_BT_2}\right] \quad (38)$$

where $\omega^\pm = \omega \pm \Omega$. Quite recently, equations (37), (38) were also obtained in [21].

In the case where the rotation axis in directed along the x-axis, we obtain [18]

$$F_z = -\frac{3\hbar}{32\pi z_0^4} \int_{-\infty}^{+\infty} d\omega \left\{ \begin{array}{l} \left[\Delta'(\omega)\alpha''(\omega)\coth\frac{\hbar\omega}{2k_BT_1} + \Delta''(\omega)\alpha'(\omega)\coth\frac{\hbar\omega}{2k_BT_2}\right] + \\ + 3\left[\Delta'(\omega)\alpha''(\omega^+)\coth\frac{\hbar\omega^+}{2k_BT_1} + \Delta''(\omega)\alpha'(\omega^+)\coth\frac{\hbar\omega}{2k_BT_2}\right] \end{array} \right\} \quad (39)$$

$$\dot{Q} = \frac{\hbar}{16\pi z_0^3} \int_{-\infty}^{+\infty} d\omega\,\omega \left\{ \begin{array}{l} \Delta''(\omega)\alpha''(\omega)\left[\coth\frac{\hbar\omega}{2k_BT_2} - \coth\frac{\hbar\omega}{2k_BT_1}\right] + \\ + 3\Delta''(\omega)\alpha''(\omega^+)\left[\coth\frac{\hbar\omega}{2k_BT_2} - \coth\frac{\hbar\omega^+}{2k_BT_1}\right] \end{array} \right\} \quad (40)$$

$$M_x = \frac{3\hbar}{16\pi z_0^3} \int_{-\infty}^{+\infty} d\omega\,\Delta''(\omega)\alpha''(\omega^+)\left[\coth\frac{\hbar\omega^+}{2k_BT_1} - \coth\frac{\hbar\omega}{2k_BT_2}\right] \quad (41)$$

### 4.4 Two rotating particles in a vacuum background

Figure 4 shows a configuration (rotation around the z-axis) and coordinate systems used $\Sigma$ (resting particle 2) and $\Sigma'$ (rotating particle 1, co-rotating reference frame).

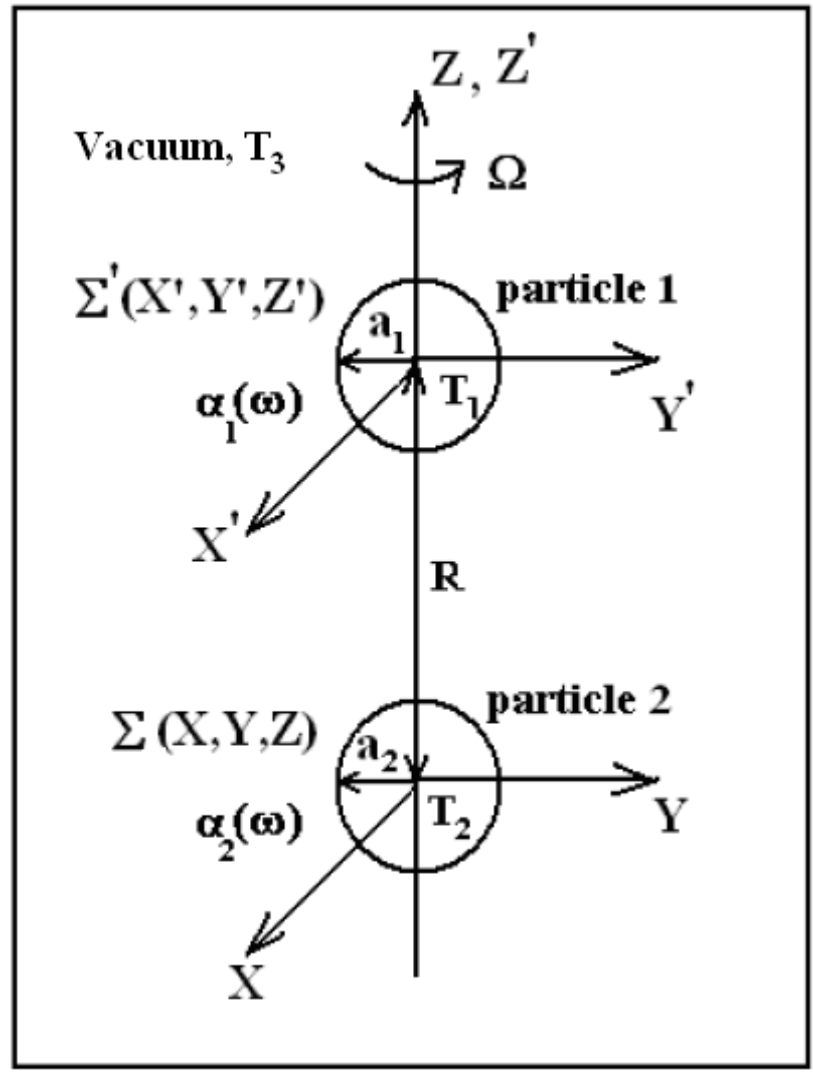

Fig. 4. Two rotating particles

The problem has been solved in relativistic statement with account of retardation, assuming the particles to be nonmagnetic [14]. The resulting expressions for $U(R), \dot{Q}$ and $M_z$ are given by

$$U(R) = -\frac{\hbar}{2\pi} \int_0^\infty d\omega \left(\frac{\omega^2}{\hbar c^2}\right)^2 \cdot$$

$$\begin{cases}
\mathrm{Re}[D_{iz}(\omega,\mathbf{R})D_{iz}(\omega,\mathbf{R})\alpha_2(\omega)]\alpha_1''(\omega)\coth\frac{\hbar\omega}{2k_BT_1} + \frac{1}{2}\mathrm{Re}[D_{i\mu}(\omega,\mathbf{R})D_{i\mu}(\omega,\mathbf{R})\alpha_2(\omega)] \cdot \\
\cdot \left[\alpha_1''(\omega_+)\coth\frac{\hbar\omega_+}{2k_BT_1} + \alpha_1''(\omega_-)\coth\frac{\hbar\omega_-}{2k_BT_1}\right] + \\
+ \mathrm{Re}[D^*_{iz}(\omega,\mathbf{R})D_{iz}(\omega,\mathbf{R})\alpha_1(\omega)]\alpha_2''(\omega)\coth\frac{\hbar\omega}{2k_BT_2} + \\
\frac{1}{2}\mathrm{Re}[D^*_{i\mu}(\omega,\mathbf{R})D_{i\mu}(\omega,\mathbf{R})(\alpha_1(\omega_+)+\alpha_1(\omega_-))] \cdot \alpha_2''(\omega)\coth\frac{\hbar\omega}{2k_BT_2} + \\
+ \mathrm{Re}[\alpha_1^*(\omega)\alpha_2(\omega)D_{iz}(\omega,\mathbf{R})]D_{iz}''(\omega,\mathbf{R})\coth\frac{\hbar\omega}{2k_BT_3} + \\
+ \frac{1}{2}\mathrm{Re}[(\alpha_1^*(\omega_+)+\alpha_1^*(\omega_-))\alpha_2(\omega)D_{i\mu}(\omega,\mathbf{R})] \cdot D_{i\mu}''(\omega,\mathbf{R})\coth\frac{\hbar\omega}{2k_BT_3} + \\
+ \mathrm{Re}[\alpha_1(\omega)\alpha_2(\omega)D_{iz}(\omega,\mathbf{R})]D_{iz}''(\omega,\mathbf{R})\coth\frac{\hbar\omega}{2k_BT_3} + \\
+ \frac{1}{2}\mathrm{Re}[(\alpha_1(\omega_+)+\alpha_1(\omega_-))\alpha_2(\omega)D_{i\mu}(\omega,\mathbf{R})] \cdot D_{i\mu}''(\omega,\mathbf{R})\coth\frac{\hbar\omega}{2k_BT_3}
\end{cases} \quad (42)$$

$$\dot{Q} = -\frac{2\hbar}{3\pi c^3}\int_0^\infty d\omega \omega^4 \begin{Bmatrix} \alpha_1''(\omega)\left[\coth\frac{\hbar\omega}{2k_B T_1} - \coth\frac{\hbar\omega}{2k_B T_3}\right] + \\ \alpha_1''(\omega_+)\left[\coth\frac{\hbar\omega_+}{2k_B T_1} - \coth\frac{\hbar\omega}{2k_B T_3}\right] + \\ \alpha_1''(\omega_-)\left[\coth\frac{\hbar\omega_-}{2k_B T_1} - \coth\frac{\hbar\omega}{2k_B T_3}\right] \end{Bmatrix} + \left\{-\frac{\hbar}{\pi}\int_0^\infty d\omega\omega\left(\frac{\omega^2}{\hbar c^2}\right)\cdot\right.$$

$$\cdot\begin{Bmatrix} \text{Im}[D_{iz}(\omega,\mathbf{R})D_{iz}(\omega,\mathbf{R})\alpha_2(\omega)]\alpha_1''(\omega)\coth\frac{\hbar\omega}{2k_B T_1} + \frac{1}{2}\text{Im}[D_{i\mu}(\omega,\mathbf{R})D_{i\mu}(\omega,\mathbf{R})\alpha_2(\omega)]\cdot \\ \cdot\left[\alpha_1''(\omega_+)\coth\frac{\hbar\omega_+}{2k_B T_1} + \alpha_1''(\omega_-)\coth\frac{\hbar\omega_-}{2k_B T_1}\right] - \\ -\text{Im}[D^*_{iz}(\omega,\mathbf{R})D_{iz}(\omega,\mathbf{R})\alpha_1(\omega)]\alpha_2''(\omega)\coth\frac{\hbar\omega}{2k_B T_2} - \\ -\frac{1}{2}\text{Im}[D^*_{i\mu}(\omega,\mathbf{R})D_{i\mu}(\omega,\mathbf{R})(\alpha_1(\omega_+) + \alpha_1(\omega_-))]\cdot\alpha_2''(\omega)\coth\frac{\hbar\omega}{2k_B T_2} + \\ +\text{Im}[\alpha_1^*(\omega)\alpha_2(\omega)D_{iz}(\omega,\mathbf{R})]D_{iz}''(\omega,\mathbf{R})\coth\frac{\hbar\omega}{2k_B T_3} + \\ +\frac{1}{2}\text{Im}[(\alpha_1^*(\omega_+) + \alpha_1^*(\omega_-))\alpha_2(\omega)D_{i\mu}(\omega,\mathbf{R})]\cdot D_{i\mu}''(\omega,\mathbf{R})\coth\frac{\hbar\omega}{2k_B T_3} - \\ -\text{Im}[\alpha_1(\omega)\alpha_2(\omega)D_{iz}(\omega,\mathbf{R})]D_{iz}''(\omega,\mathbf{R})\coth\frac{\hbar\omega}{2k_B T_3} - \\ -\frac{1}{2}\text{Im}[(\alpha_1(\omega_+) + \alpha_1(\omega_-))\alpha_2(\omega)D_{i\mu}(\omega,\mathbf{R})]\cdot D_{i\mu}''(\omega,\mathbf{R})\coth\frac{\hbar\omega}{2k_B T_3} \end{Bmatrix} \quad (43)$$

$$M_z = -\frac{2\hbar}{3\pi c^3}\int_0^\infty d\omega\omega^3\begin{Bmatrix} \alpha_1''(\omega_-)\left[\coth\frac{\hbar\omega}{2k_B T_1} - \coth\frac{\hbar\omega}{2k_B T_3}\right] - \\ -\alpha_1''(\omega_+)\left[\coth\frac{\hbar\omega_+}{2k_B T_1} - \coth\frac{\hbar\omega}{2k_B T_3}\right] \end{Bmatrix} - \frac{\hbar}{2\pi}\int_0^\infty d\omega\left(\frac{\omega^2}{\hbar c^2}\right)^2\cdot$$

$$\cdot\begin{Bmatrix} \text{Im}[(D^2_{xx}(\omega,\mathbf{R}) + D^2_{yy}(\omega,\mathbf{R}))\alpha_2(\omega)]\cdot\left(\alpha_1''(\omega_-)\coth\frac{\hbar\omega_-}{2k_B T_1} - \alpha_1''(\omega_+)\coth\frac{\hbar\omega_+}{2k_B T_1}\right) + \\ +\text{Im}[(|D_{xx}(\omega,\mathbf{R})|^2 + |D_{yy}(\omega,\mathbf{R})|^2)\alpha_2(\omega)]\cdot(\alpha_1''(\omega_+) - \alpha_1''(\omega_-))\coth\frac{\hbar\omega}{2k_B T_2} + \\ +[2\text{Re}(\alpha_2(\omega)D_{xx}(\omega,\mathbf{R}))\text{Im}\,D_{xx}(\omega,\mathbf{R}) + 2\text{Re}(\alpha_2(\omega)D_{yy}(\omega,\mathbf{R}))\text{Im}\,D_{yy}(\omega,\mathbf{R})]\cdot \\ \cdot(\alpha_1''(\omega_+) - \alpha_1''(\omega_-))\coth\frac{\hbar\omega}{2k_B T_3} \end{Bmatrix} \quad (44)$$

$$D_{ik}(\omega,\mathbf{R}) = \left(-\frac{\omega^2}{\hbar c^2}\right)\exp(i\omega R/c)\left[\left(\frac{\omega^2}{c^2 R} + \frac{i\omega}{cR} - \frac{1}{R^3}\right)(\delta_{ik} - n_i n_k) + 2\left(\frac{1}{R^3} - \frac{i\omega}{cR^2}\right)n_i n_k\right] \quad (45)$$

where $\mathbf{n} = \mathbf{R}/R$, $i,k = x,y,z$; $\mu = x,y$. In the nonrelativistic limit $c \to \infty$, Eqs. (43)—(45) are reduced to

$$U = -\frac{\hbar}{2\pi R^6} \int_0^\infty d\omega \left[ 4\left( \alpha_1''(\omega)\alpha_2'(\omega) \coth\frac{\hbar\omega}{2k_B T_1} + \alpha_1'(\omega)\alpha_2''(\omega) \coth\frac{\hbar\omega}{2k_B T_2} \right) + \right.$$
$$+ \alpha_1''(\omega_+)\alpha_2'(\omega) \coth\frac{\hbar\omega_+}{2k_B T_1} + \alpha_1'(\omega_+)\alpha_2''(\omega) \coth\frac{\hbar\omega}{2k_B T_2} +$$
$$\left. + \alpha_1''(\omega_-)\alpha_2'(\omega) \coth\frac{\hbar\omega_-}{2k_B T_1} + \alpha_1'(\omega_-)\alpha_2''(\omega) \coth\frac{\hbar\omega}{2k_B T_2} \right] \quad (46)$$

$$\dot{Q} = \frac{\hbar}{\pi R^6} \int_0^\infty d\omega\, \omega\, \alpha_2''(\omega) \left[ 4\alpha_1''(\omega)\left( \coth\frac{\hbar\omega}{2k_B T_2} - \coth\frac{\hbar\omega}{2k_B T_1} \right) + \right.$$
$$\left. + \alpha_1''(\omega_+)\left( \coth\frac{\hbar\omega}{2k_B T_2} - \coth\frac{\hbar\omega_+}{2k_B T_1} \right) + \alpha_1''(\omega_-)\left( \coth\frac{\hbar\omega}{2k_B T_2} - \coth\frac{\hbar\omega_-}{2k_B T_1} \right) \right] \quad (47)$$

$$M_z = -\frac{\hbar}{\pi R^6} \int_0^\infty d\omega\, \alpha_2''(\omega) \left[ \begin{array}{l} \alpha_1''(\omega_-)\left( \coth\frac{\hbar\omega_-}{2k_B T_1} - \coth\frac{\hbar\omega}{2k_B T_2} \right) - \\ -\alpha_1''(\omega_+)\left( \coth\frac{\hbar\omega_+}{2k_B T_1} - \coth\frac{\hbar\omega}{2k_B T_2} \right) \end{array} \right] \quad (48)$$

At $\Omega = 0$, Eq. (47) was first obtained in [22]. In the case $\mathbf{\Omega} = (\Omega,0,0)$ (rotation around the $x$-axis), the corresponding formulas differ from (42)—(44) by the cyclic transposition of $x, y, z$. Formulas (46)—(48) have a very close structure with minor differences of numerical factors [14]. In the particular case $\Omega = 0$, Eq. (42) takes the form [14]

$$U(R) = -\frac{\hbar}{2\pi} \int_0^\infty d\omega \left( \frac{\omega^2}{\hbar c^2} \right)^2 \cdot$$
$$\cdot \left\{ \begin{array}{l} \text{Re}[D_{ik}(\omega,\mathbf{R})D_{ik}(\omega,\mathbf{R})]\text{Im}[\alpha_1(\omega)\alpha_2(\omega)]\coth\dfrac{\hbar\omega}{2k_B T_1} + \\ + \text{Im}[D_{ik}(\omega,\mathbf{R})D_{ik}(\omega,\mathbf{R})]\text{Re}[\alpha_1(\omega)\alpha_2(\omega)]\coth\dfrac{\hbar\omega}{2k_B T_3} + \\ + 2 D_{ik}'(\omega,\mathbf{R})D_{ik}''(\omega,\mathbf{R})\alpha_1''(\omega)\alpha_2''(\omega)\left[ \coth\dfrac{\hbar\omega}{2k_B T_3} - \coth\dfrac{\hbar\omega}{2k_B T_1} \right] + \\ + D_{ik}'(\omega,\mathbf{R})D_{ik}'(\omega,\mathbf{R})\alpha_1'(\omega)\alpha_2''(\omega)\left[ \coth\dfrac{\hbar\omega}{2k_B T_2} - \coth\dfrac{\hbar\omega}{2k_B T_1} \right] + \\ + D_{ik}''(\omega,\mathbf{R})D_{ik}''(\omega,\mathbf{R})\alpha_1'(\omega)\alpha_2''(\omega)\left[ \coth\dfrac{\hbar\omega}{2k_B T_1} + \coth\dfrac{\hbar\omega}{2k_B T_2} - 2\coth\dfrac{\hbar\omega}{2k_B T_3} \right] \end{array} \right\} \quad (49)$$

To our knowledge, Eq.(49) is new. In the case of total thermal equilibrium $T_1 = T_2 = T_3 = T$, Eq. (49) is reduced to

$$U(R) = -\frac{\hbar}{2\pi} \text{Im}\left[ \int_0^\infty d\omega \left( \frac{\omega^2}{\hbar c^2} \right)^2 \coth\frac{\hbar\omega}{2k_B T} \alpha_1(\omega)\alpha_2(\omega) D_{ik}(\omega,\mathbf{R}) D_{ik}(\omega,\mathbf{R}) \right] \quad (50)$$

From (50) we obtain the classical result by Casimir and Polder [23]

$$U(R) = -\frac{\hbar}{\pi R^6}\int_0^\infty d\omega\, \alpha_1(i\omega)\alpha_2(i\omega)\exp(-2\omega R/c)\left(\frac{\omega R}{c}\right)^4 \cdot$$
$$\cdot\left[1 + 2\left(\frac{\omega R}{c}\right)^{-1} + 5\left(\frac{\omega R}{c}\right)^{-2} + 6\left(\frac{\omega R}{c}\right)^{-3} + 3\left(\frac{\omega R}{c}\right)^{-4}\right]$$
(51)

If the particle is magnetically polarizable (without electrical polarization), all the formulas in this section are valid when replacing electric polarizabilities by magnetic ones. In the case, where both of the polarizations are present, apart from the separate contributions of different type polarization, one should also take into account the mixed terms since the each type of the particle polarization may create another type (electric –magnetic and vice versa).

### 4.5 Two parallel thick plates in relative motion

General geometrical and thermal configuration is shown in Fig. 4. To date, general relativistic problem in this case is still being debated even under total thermal equilibrium [7,10 24-28]. In [10,17], we have proposed the correspondence principle between the configurations particle-surface and surface-surface. This allowed us to find an unambiguous solution to the problem in configuration surface-surface at $V \ll c$ and $T_1 = T_2 = T_3 = T$ (indexes 1,2 numerate the moving and resting plates, respectively)

$$F_x(l) = -\frac{\hbar S}{4\pi^3}\int_0^\infty d\omega \int_{-\infty}^{+\infty} dk_x \int_{-\infty}^{+\infty} dk_y\, k_x \cdot$$
$$\cdot\left[\frac{\mathrm{Im}\,\Delta_{1e}(\omega^+)\mathrm{Im}(\exp(-2q_0 l)\Delta_{2e}(\omega))}{\left|1-\exp(-2q_0 l)\Delta_{1e}(\omega^+)\Delta_{2e}(\omega)\right|^2} + \frac{\mathrm{Im}\,\Delta_{1m}(\omega^+)\mathrm{Im}(\exp(-2q_0 l)\Delta_{2m}(\omega))}{\left|1-\exp(-2q_0 l)\Delta_{1m}(\omega^+)\Delta_{2m}(\omega)\right|^2}\right] \cdot$$
$$\cdot\left[\coth\left(\frac{\hbar\omega}{2k_B T}\right) - \coth\left(\frac{\hbar\omega^+}{2k_B T}\right)\right]$$
(52)

$$F_z(l) = -\frac{\hbar S}{4\pi^3}\int_0^\infty d\omega \int_{-\infty}^{+\infty} dk_x \int_{-\infty}^{+\infty} dk_y \cdot$$
$$\cdot\left[\left(\frac{\mathrm{Im}\,\Delta_{1e}(\omega^+)\mathrm{Re}(q_0\exp(-2q_0 l)\Delta_{2e}(\omega))}{\left|1-\exp(-2q_0 l)\Delta_{1e}(\omega^+)\Delta_{2e}(\omega)\right|^2}\coth\left(\frac{\hbar\omega^+}{2k_B T}\right) + \right.\right.$$
$$\left.+\frac{\mathrm{Im}\,\Delta_{1m}(\omega^+)\mathrm{Re}(q_0\exp(-2q_0 l)\Delta_{2m}(\omega))}{\left|1-\exp(-2q_0 l)\Delta_{1m}(\omega^+)\Delta_{2m}(\omega)\right|^2}\coth\left(\frac{\hbar\omega^+}{2k_B T}\right)\right) +$$
$$+\left(\frac{\mathrm{Re}\,\Delta_{1e}(\omega^+)\mathrm{Im}(q_0\exp(-2q_0 l)\Delta_{2e}(\omega))}{\left|1-\exp(-2q_0 l)\Delta_{1e}(\omega^+)\Delta_{2e}(\omega)\right|^2}\coth\left(\frac{\hbar\omega}{2k_B T}\right) + \right.$$
$$\left.\left.+\frac{\mathrm{Re}\,\Delta_{1m}(\omega^+)\mathrm{Im}(q_0\exp(-2q_0 l)\Delta_{2m}(\omega))}{\left|1-\exp(-2q_0 l)\Delta_{1m}(\omega^+)\Delta_{2m}(\omega)\right|^2}\coth\left(\frac{\hbar\omega}{2k_B T}\right)\right)\right]$$
(53)

$$\dot{Q}(l) = \frac{\hbar S}{4\pi^3} \int_0^\infty d\omega \int_{-\infty}^{+\infty} dk_x \int_{-\infty}^{+\infty} dk_y \, \omega^+ \cdot$$

$$\cdot \left[ \frac{\operatorname{Im}\Delta_{1e}(\omega^+)\operatorname{Im}(\exp(-2q_0 l)\Delta_{2e}(\omega))}{\left|1-\exp(-2q_0 l)\Delta_{1e}(\omega^+)\Delta_{2e}(\omega)\right|^2} + \frac{\operatorname{Im}\Delta_{1m}(\omega^+)\operatorname{Im}(\exp(-2q_0 l)\Delta_{2m}(\omega))}{\left|1-\exp(-2q_0 l)\Delta_{1m}(\omega^+)\Delta_{2m}(\omega)\right|^2} \right] \cdot \qquad (54)$$

$$\cdot \left[ \coth\left(\frac{\hbar\omega}{2k_B T}\right) - \coth\left(\frac{\hbar\omega^+}{2k_B T}\right) \right]$$

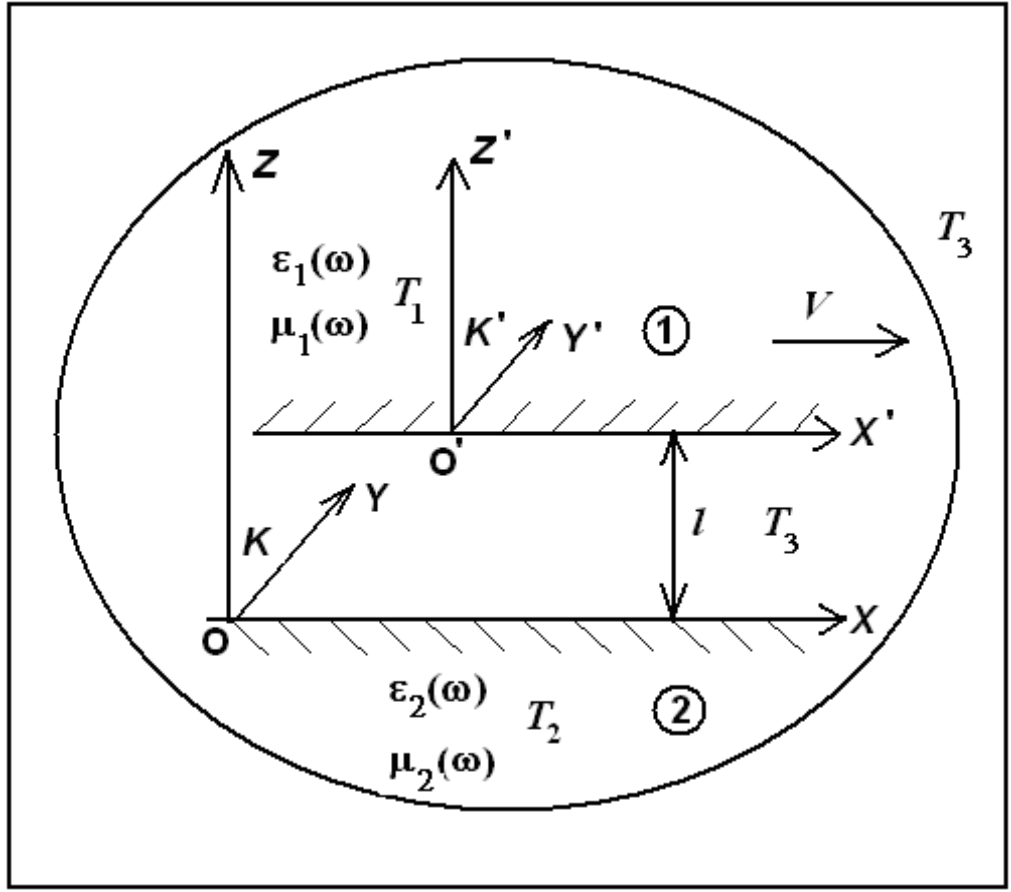

Fig. 5. Configuration of the two thick plates embedded in vacuum background

In the case $T_1 \neq T_2$, and at an arbitrary temperature $T_3$, the contributions to integrals (52)—(54) associated with the evanescent electromagnetic modes $k > \omega/c$ are also correct. Therefore, in the complete nonrelativistic limit $c \to \infty$ formulas (52)—(54) will adequately describe the interaction in thermal disequilibrium if we replace $\omega/T$ by $\omega/T_2$, and $\omega^+/T$ by $\omega^+/T_1$. The resulting expressions have the form

$$F_x(l) = -\frac{\hbar S}{4\pi^3}\int_0^\infty d\omega \int_{-\infty}^{+\infty} dk_x \int_{-\infty}^{+\infty} dk_y k_x \frac{\exp(-2kl)}{\left|1-\exp(-2kl)\Delta_1(\omega^+)\Delta_2(\omega)\right|^2}\Delta_1''(\omega^+)\Delta_2''(\omega)\cdot$$
$$\cdot\left[\coth(\hbar\omega/2k_B T_2) - \coth(\hbar\omega^+/2k_B T_1)\right] + (...) \tag{55}$$

$$F_z(l) = -\frac{\hbar S}{4\pi^3}\int_0^\infty d\omega \int_{-\infty}^{+\infty} dk_x \int_{-\infty}^{+\infty} dk_y k \frac{\exp(-2kl)}{\left|1-\exp(-2kl)\Delta_1(\omega^+)\Delta_2(\omega)\right|^2} \cdot \quad + (...)$$
$$\cdot\left[\Delta_1''(\omega^+)\Delta_2'(\omega)\coth(\hbar\omega^+/2k_B T_1) + \Delta_1'(\omega^+)\Delta_2''(\omega)\coth(\hbar\omega/2k_B T_2)\right] \tag{56}$$

$$\dot{Q}(l) = \frac{\hbar S}{4\pi^3}\int_0^\infty d\omega \int_{-\infty}^{+\infty} dk_x \int_{-\infty}^{+\infty} dk_y \omega^+ \frac{\exp(-2kl)}{\left|1-\exp(-2kl)\Delta_1(\omega^+)\Delta_2(\omega)\right|^2}\Delta_1''(\omega^+)\Delta_2''(\omega) + (...)\cdot$$
$$\cdot\left[\coth(\hbar\omega/2k_B T_2) - \coth(\hbar\omega^+/2k_B T_1)\right] \tag{57}$$

The terms (...) in (55)—(57) have the same structure making the change
$$\frac{\varepsilon_{1,2}(\omega)-1}{\varepsilon_{1,2}(\omega)+1} \to \frac{\mu_{1,2}(\omega)-1}{\mu_{1,2}(\omega)+1}.$$

It turns out that only formula (55) agrees with [7], while the corresponding expressions for $F_z$ and $\dot{Q}$ in [7] principally differ from (56),(57) and turn out to be incorrect, as we have shown in detail [10,17]. The results [29] for the configuration particle –surface are also wrong [10,11,17]. In the static case $V=0$ out of thermal equilibrium $T_1 \neq T_2 \neq T_3$, the problem has been solved in [2]. We refer the reader to the corresponding results for $F_z$ and $\dot{Q}$ in that work.

## 5. Concluding remarks

The volume of this paper does not allow us to illustrate the obtained results numerically, since this requires much more space. Nevertheless, the analytical expressions that we have listed, can serve for future theoretical elaboration and practical applications. They demonstrate an overall consistency with the vast number of works of other authors and contain the well-known results as the corresponding limiting cases. All the problems have been solved from first principles, in full compliance with the principle of relativistic invariance and fluctuation electrodynamics. We should stress, once again, close resemblance between the systems with thermal and dynamic disequilibrium. This is clearly visible for all of the formulas obtained, since the frequency, temperature, and velocity-dependent factors of the first body, corresponding to the moving/rotating particle or a particle in thermal disequilibrium with other bodies (vacuum background, plane surface, another particle) are combined into a single variable, indicating the type of disequilibrium.